# Emotion Detection Using Noninvasive Low Cost Sensors

Daniela Girardi, Filippo Lanubile, Nicole Novielli
University of Bari 'Aldo Moro'
d.girardi2@studenti.uniba.it; {filippo.lanubile,nicole.novielli}@uniba.it

*Abstract*—Emotion recognition from biometrics is relevant to a wide range of application domains, including healthcare. Existing approaches usually adopt multi-electrodes sensors that could be expensive or uncomfortable to be used in real-life situations. In this study, we investigate whether we can reliably recognize high vs. low emotional valence and arousal by relying on noninvasive low cost EEG, EMG, and GSR sensors. We report the results of an empirical study involving 19 subjects. We achieve state-of-the-art classification performance for both valence and arousal even in a cross-subject classification setting, which eliminates the need for individual training and tuning of classification models.

## 1. Introduction

Emotion recognition from biometrics is a consolidate research field [4][8][9][10][13][19][20][27][28][29] relevant to a wide range of application domains, from human-computer interaction [18] to software engineering [7], and healthcare [1]. Brain-related measures, such as electroencephalography (EEG) and skin-related measures, such as galvanic skin response (GSR) and electromyography (EMG) are among the most popular and widely adopted physiological measures for affect detection [18], also in combination with heart-rate, blood volume, and respiration metrics [11][14].

Recent approaches to emotion recognition based on biofeedback successfully adopt sensors with multiple channels for signal recording. It is the case, for example, of EEG helmets using from 32 to 60 electrodes for capturing the brain activities [14][19][27]. As for EMG, sensors are usually placed over the face [14], since facial muscles clearly reflect changes in emotional states [6]. Biomedical toolboxes are also employed for capturing psychological measures from multiple sensors [10].

Even if successfully employed for emotion recognition, such sensors might be expensive or uncomfortable to use in real-life situations. It is the case, for example of the scenario of our ongoing research whose long-term goal is to detect the emotional state of patients with impaired cognition during their medical treatments. In such a scenario, noninvasiveness, comfort and ease of use for both the therapist and the patient are crucial criteria in the choice of the devices to employ for biofeedback acquisition. Sensors should be comfortably worn by patient without causing additional distress or negative emotions. Also, it becomes important to rely on low cost devices to allow large-scale adoption of affect-aware approaches to medical treatments.

Consistently with the long-term goals of our research, in the present study we investigate the suitability of noninvasive low cost sensors for emotion recognition. Specifically, we propose a partial replication of the study described in [14] for the recognition of emotional valence and arousal [24]. We formulate our research question as follows:

*RQ: Can we acquire physiological measures from noninvasive, low cost EEG, GSR, and EMG sensors to accurately predict emotions?*

We investigate what are the most relevant physiological measures for both valence and arousal to define the optimal combination of sensors for the two recognition tasks. Furthermore, we investigate whether it is possible to reliably identify emotions through models trained in a cross-subject classification setting. This is particularly important in our research since individual training of classification model is practically impossible with patients with impaired cognition and mobility, which are the main actors of our target scenario. Both the cross-subject setting and the use of noninvasive low cost sensors represent the main novelty with respect to the original study by Koelstra et al. [14] that we partially replicate.

The paper is structured as follows. In Section 2 we describe the theoretical background and offer a survey on recent related work on emotion detection based on biometrics. In Section 3 we describe the empirical study, including a detailed description of the sensors used, the video selected for emotion elicitation, and the experimental protocol adopted. Finally, we report our classification experiment in Section 4 and discuss results in Section 5. Conclusions and directions for future work are reported in Section 6.

## 2. Background and Related Work

### 2.1. Circumplex Model of Affect

Psychologists worked at decoding emotions for decades. by focusing on how classify emotions, what is their functioning, and what is the role played by cognition in their triggering [3]. Two points of view prevail. The first one assumes that a set of discrete emotions exists [5][17]. Conversely, the second one consider emotions as a continuous function of one or more dimensions. It is the case, for example, of the such as Plutchik's emotion wheel [22] or the 'Circumplex Model' of affect by Russel [24]. Consistently with the original study we intend to partially replicate [14], we refer to the Russell's model, which represents emotions along a bi-dimensional representation schema, including valence and arousal in the horizontal and vertical axes, respectively.

*Valence* describes the pleasantness of an emotional state. Pleasant emotional states, such as joy or amusement, are associated to high valence, while unpleasant ones, such as sadness or fear, are associated to low valence. *Arousal* describes the level of activation of the emotional state ranging from inactive (low arousal), as in boredom or sleepiness, to active (high arousal) as in excitement or anger.

## 2.2. Sensing Emotions from Biometrics

Affective computing studies have largely investigated emotion recognition from several physiological signals, either alone or in combination [4][8][10][13][19][20][27][28][29], thus confirming the link between emotion and physiological feedback [2]. In the following we describe the physiological measures used in the present study, which we selected consistently with our long-term goal of supporting the wellbeing of people with impaired movements or damaged cognition, during their medical treatment and physiotherapy. As such, our emotion recognition approach exploits only on sensors that can be reliably used during the medical treatments.

**Electroencephalography (EEG)** consists in the recording of the brain-activity, which can be captured using electrodes placed on the surface of the scalp or in the forehead. The link between variations in specific frequency bands, i.e. *alpha* (4-12,5 Hz), *beta* (13-30 Hz), *gamma* (>30 Hz), *delta* (<4 Hz), and *theta* (4-7,5 Hz), and cognitive [7] as well as emotional states [19][27] has been recognized.

**Galvanic Skin Response (GSR)** is a measure of skin conductance, that is, of the electrical activity of the skin due to the variation in human body sweating. The GSR signal consists of two main components, namely the *tonic* and the *phasic* signals. The tonic component indicates the basic level of skin conductance, which varies from person to person, while the phasic component changes according to specific external stimuli such as sounds, noises, changes in light condition, etc. [25]. Studies in psychology show how GSR varies considerably with respect to changes in emotional intensity, especially for emotions with high arousal [2]. As such, GSR has been widely employed in emotion recognition [4][8][10][14][20][21].

**Electromyographic signal (EMG)** describes the electrical activity of contracting muscles. Whenever a contraction occurs, electricity is generated and propagated in tissues, bones and in the nearby skin area. EMG activity is linearly related to the amount of muscle contraction and the number of muscles contracted. However, EMG activity is measurable even when no observable contractions can be seen, for example, when we control the body so that certain behaviors do not occur. This makes EMG an excellent technique to monitor the cognitive-behavioral process in addition to simple observation, as well as a predictor for emotions [4][8][10][20].

**Hearth-related measurements**, such as blood volume or hearth rate, are also employed for recognition of emotional and cognitive states [4][8][7][23]. The heart rate is usually derived by applying conversion algorithms to the signal captured by a plethysmograph, which is an optical sensor usually applied on a finger.

## 3. Empirical Study

We performed an empirical study with 19 subjects (16 males, 3 females), recruited on a voluntary basis among university students, friends, and relatives, aged between 20 and 40. Subjects were required to watch a sequence of videos while recording some physiological measures.

### 3.1. Video Selection and Preparation

The videos were selected among the 40 annotated music videos included in the DEAP dataset [14]. Each video in the DEAP is associated with valence and arousal values, on a scale from 1 to 9. Consistently to [14], four classes were identified, corresponding to the four quadrants of the emotional space in the Circumplex Model of Affect. We discretize the scores by associating a Low Valence (LV) or Low Arousal (LA) label to values lower or equal to 4.5. Conversely, we discretize as High Valence (HV) or High Arousal (HA) those scores higher or equal to 6. As a result, we have four classes of emotions corresponding to the four quadrants of the bi-dimensional model of emotions. TABLE I. shows the ranges for Valence and Arousal scores associated to the videos included in each class of our experiment. The original ID of selected videos ('Online ID' in the DEAP database [14]) is also reported in the fourth column.

TABLE I. RANGES FOR AROUSAL AND VALENCE SCORES FOR THE VIDEOS IN THE FOUR EMOTION CLASSES.

| Emotion Class | Arousal | Valence | Video IDs |
|---|---|---|---|
| LAHV | [ 3,86 - 4,21 ] | [ 6,57 – 7,13 ] | 24, 80 |
| LALV | [ 2,75 - 2,93 ] | [ 3,25 - 3,33 ] | 41, 96 |
| HAHV | [ 6,40 -7,33 ] | [ 7,07 - 7,20 ] | 63, 88 |
| HALV | [ 6,13- 6,33 ] | [ 3,53 - 3,93 ] | 56, 111 |

### 3.2. Biometric Sensors

EEG raw signals were recorded with the BrainLink[1] headset (see Figure 1.a) connected via Bluethooth to a dedicated recording laptop (Intel i7, 2.5 GHz), using the Neuroview acquisition software. EEG was recorded at a sampling rate of 512 Hz. The sensor is equipped with three metal sensors, two on the forehead and one on the left earlobe. One of the two sensors placed in the forehead is the active electrode capturing raw brainwaves signals. The second one captures the noise generated by intentional and unintentional body and head movements. The signal is computed as the potential difference between the signal captured by the active electrode and the reference electrode placed on the earlobe.

---

[1] www.mindtecstore.com/en/brainlink

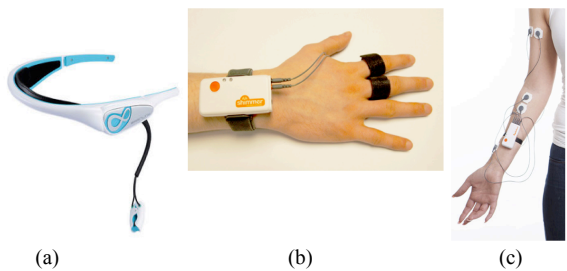

Figure. 1. Biometric sensors: (a) EEG, (b) GSR, and (c) EMG.

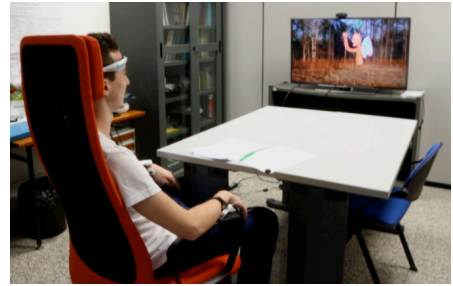

Figure. 2. A participant during the experiment.

GSR is captured at a sampling rate of 128 Hz using the Shimmer GSR+Unit[2], a device that measures the galvanic response of the skin through two electrodes positioned on the palmar surface of the first phalanx of two different fingers (see Figure 1.b). The device also allows capturing the hearth rate through a plethysmograph placed on a third finger. We captured the hearth rate during the experiments but could not used it due to poor quality of the signal recorded.

EMG is captured at a sampling rate of 512 Hz using Shimmer EMG[3], a device that records the electrical activity associated with skeletal muscle contraction. The EMG data are measured by two channels, each with a positive and a negative electrode. A fifth electrode acts as a reference. Positive and negative electrodes should be positioned parallel to the fibers of the muscle that you want to measure, near the center of the muscle. The reference electrode must instead be positioned in a neutral point in the body, e.g. on the wrist bone. Figure 1.c shows an example of the arrangement of the five electrodes. Both Shimmer devices are connected to the laptop via Bluetooth. Data are collected and recorded using the ConsensysPRO software.

### 3.3. Experimental Protocol

The experimental sessions were conducted in the same laboratory with controlled illumination, during March 2017. The video sequence was displayed on a 42" screen at a full HD resolution, filling the full screen. The screen had embedded speakers and the volume was set to a relatively loud level to guarantee the comfort of participants. To minimize head and body movements that could introduce noise in the raw signals, subjects were seating on a chair with comfortable backrest and armrest. The chair was positioned at one meter from the screen, as shown in Figure 2.

Each session lasted 30 minutes overall. The log of the collected biometrics was enriched with timestamp and the experimenter (i.e., the first author) took note of the exact system time in the moment the session started, to allow synchronous analysis of signals with respect to the videos. Prior to the experiment, each subject signed a consent form. Next, the experimenter welcomed the subjects in the lab and shortly explained the goal of our ongoing research as well as the specific purpose of the experiment. During the initial setup phase, the experimenter was available to answer any questions. Before starting the signal recording, the experimenter applied all sensors and verified that they were properly capturing and recording the physiological signals. Sensors were connected to a laptop via Bluetooth and the experimenter constantly monitored their correct functioning through the interface of the signal acquisition program.

Next, the experiment started and the eight videos were presented in 4 trials, as shown in Figure 3. Each trial consists of the following steps: (i) A 30-second baseline video showing a quiet image with relaxing musing in the background; (ii) a 2-minute display of the selected music videos (1 minute per video). At the beginning of each trial, a 3-second screen displayed the current trial number to make the participant aware of her progress. We defined the visualization sequence by showing videos with increasing arousal value. This decision is motivated by the fact that increasing arousal values are associated to higher arousal levels with respect to an individual baseline values [28].

### 4. Classification Experiment

To investigate whether it is possible to reliably distinguish between high vs. low levels of arousal and valence, we built two binary classifiers, the first for arousal and the second for valence, by considering all possible combination of sensors. Hence, we performed a classification study using the data collected for all subjects involved in our experiment. For each of our 19 participants we recorded the biometrics associated to all eight videos, for a total of 152 instances overall.

Signals acquired with wearable devices are noisy and usually contains invalid data due to factors such as the loss of contact between the electrodes and skin, eye blink, or movement artifacts [28]. As such, the raw signals acquired during the experimental sessions should be cleaned to allow reliable analysis of data. We followed the approach proposed by Fritz et al. [7] to extract the alpha, beta, gamma, delta, and theta frequencies from the EEG signal and to derive the phasic component from the GSR signal[4]. As for EMG, we filtered the signal with a band pass frequency of 20–125 HZ, as suggested by Kutz [16]. Furthermore, we performed a transaction known as *integration* [15], already adopted in previous research [8], to quantify the excitation level in the EMG signal.

---

[2] www.shimmersensing.com/products/shimmer3-wireless-gsr-sensor

[3] http://www.shimmersensing.com/products/shimmer3-emg-unit

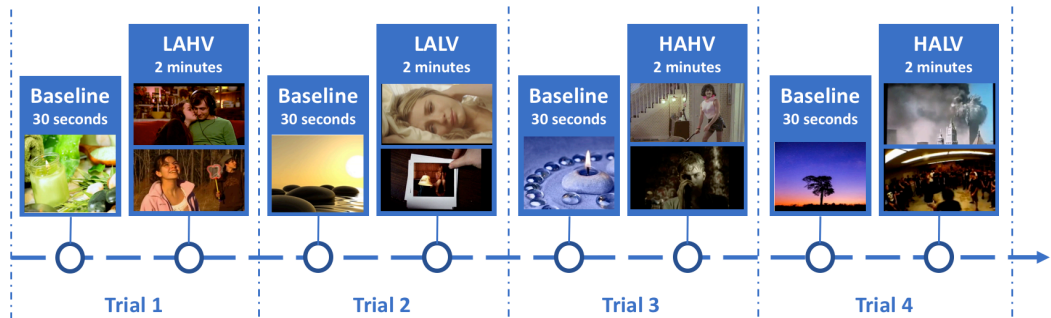

Figure 3. Timeline of the experiment.

Since every person has a unique power spectrum distribution, we correct stimulus-unrelated variations over time for all sensors by subtracting the mean value for the signal while the subject was watching the baseline videos associated to neutral emotional states. This approach was already adopted in previous studies to ensure robustness and generality of our measurements [14].

Next, we extracted the features from the last 30 seconds of each video to be used for the classifier, inspired by previous studies [7][14]. Statistical descriptive features (i.e., mean, median, standard deviation, minimum, and maximum values) are computed for all EEG, GSR, and EMG measures. For EEG, these metrics are extracted for alpha, beta, gamma, theta, delta waves as well as for the 1 Hz Attention and Meditation signals computed by the NeuroSky EEG sensor. For GSR, further features are extracted from the derivative of the phasic signal, after applying the Wavelet transform to further remove noise in the long-rang low frequency signal [7]. Since changes in electrical activity due to external stimuli are visible as peaks, we compute additional features related to the amplitude and frequency of the GSR signal. We extracted **58** features overall: 35 for EEG, 13 for GSR and 10 for EMG. The full list of features for each sensor is reported in TABLE II.

We ran our classification experiment using Weka[5], a Java library for machine learning. We trained two binary classifiers with class values in {High, Low} for both arousal and valence. As gold standard, we used the mapping to high/low labels of the original DEAP scores for each video (see TABLE I. ). We compare the performance of three popular machine-learning approaches, which have been found to be the most successful in similar contexts [7][14], namely Naïve Bayes (shortly *NB*), the Weka SMO implementation of Support Vector Machines (*SVM*) with polynomial kernel, and *J48*, an algorithm based on decision trees.

TABLE II. FEATURES EXTRACTED FROM SIGNALS.

| Signal | Extracted Features |
|---|---|
| EEG | *For the alpha, beta, gamma, delta, theta, attention, and meditation*: mean, minimum, maximum, variance, standard deviation. |
| GSR | *On the phasic component*: mean, minimum, maximum, variance, standard deviation. *On the corrected phasic component after Wavelet filter*: mean of the derivatives, average of the derivatives of negative values, and the proportion of negative values. *Considering the peaks*: mean, minimum and maximum width, ratio between number of peaks and minimum width, ratio between sum of peaks and minimum width. |
| EMG | *On the integrated signal from both channels*: mean, minimum, maximum, variance, standard deviation. |

In the original DEAP study, classifiers were trained separately for each subject involved in the experiment. However, such evaluation setting would not be consistent with the final usage scenario of our classifiers, i.e. recognition of emotions experienced by people with impaired cognition and limited mobility and vision, for which a dedicated training is unfeasible. Thus, we evaluate the performance of each classifier in a cross-subject setting, implementing a leave-one-out validation. At each iteration, one single instance of the dataset is considered for test while the remaining instances are used for training. As such, at each iteration the model is trained using all videos for all the subjects in our dataset.

## 5. Results and Discussion

In TABLE III we report the performance of the two classifiers on the entire dataset for all possible combinations of the features extracted from the three sensors. Precision, Recall, and F1 are first evaluated locally for each class (i.e. High vs. Low Arousal and High vs. Low Valence) and then globally by averaging over the performance of the classes, following the macroaveraging approach [26] implemented in Weka. The best performance in terms of F1 absolute value is highlighted in grey. The optimal settings (best F1 and statistically significant differences with respect to other settings) are highlighted in bold. We compare the differences among the various feature settings using the McNemar test, as implemented by the Caret package for R [12].

---
[4] Code available at: https://github.com/BioStack/Sensors101

TABLE III. PERFORMANCE OF CLASSIFIERS TRAINED BY ENABLING DIFFERENT FEATURE SETTINGS.

| Signals | Arousal | | | | Valence | | | |
|---|---|---|---|---|---|---|---|---|
| | Classifier | Precision | Recall | F1 | Classifier | Precision | Recall | F1 |
| Single Sensors | | | | | | | | |
| EEG | SVM | 0.605 | 0.605 | 0.605 | **SVM** | **0.567** | **0.566** | **0.563** |
| GSR | J48 | 0.671 | 0.645 | 0.630 | NB | 0.585 | 0.507 | 0.359 |
| EMG | NB | 0.315 | 0.316 | 0.315 | J48 | 0.748 | 0.599 | 0.527 |
| Combined Sensors | | | | | | | | |
| EEG+GSR | **SVM** | **0.639** | **0.638** | **0.638** | SVM | 0.553 | 0.553 | 0.551 |
| GSR+EMG | J48 | 0.653 | 0.618 | 0.596 | J48 | 0.540 | 0.539 | 0.539 |
| EEG+EMG | SVM/J48 | 0.619 | 0.618 | 0.618 | SVM | 0.559 | 0.559 | 0.559 |
| All (EEG+GSR+EMG) | SVM | 0.606 | 0.605 | 0.605 | SVM | 0.586 | 0.586 | 0.585 |

TABLE IV. COMPARISON WITH THE PERFORMANCE REPORTED IN THE ORIGINAL DEAP STUDY [14].

| Signals | Study | Description | Arousal | | Valence | |
|---|---|---|---|---|---|---|
| | | | Classifier | F1 | Classifier | F1 |
| EEG | Our Study | EEG | SVM | 0.605 | SMO | 0.563 |
| | [14] | EEG | NB | 0.583 | NB | 0.563 |
| Peripheral | Our Study | GSR + EMG | J48 | 0.596 | J48 | 0.539 |
| | [14] | GSR + EMG + respiration + blood pressure + eye blinking rate | NB | 0.553 | NB | 0.608 |

For both arousal and valence, the best performance is observed when using SVM for machine learning. However, different sensors and settings show different predictive power for the two dimensions. Of the two dimensions, the F1-score is higher for arousal. For arousal, we observe that GSR is the measure with the best classification performance among the single sensor classification settings (F1= 0.630). GSR alone performs significantly better than EEG (p-value = 0.001). Further improvement in F1-score is observed by combining GSR with EEG (F1= 0.638). Even if small, this increase in F1 is statistically significant, according to the result of the McNemar test (p = 0.001). This is consistent with previous evidence provided by psychological research [2] and confirms the predictive power of GSR-based metrics for arousal. As for valence, EEG is the sensor with the best performance in the single-sensor setting (F=0.563). The small improvement in the full feature setting (F=0.585 with all sensors) can be ignored because not statistically significant at the 0.05 level (p = 0.42).

Based on our current results, the contribute of EMG is negligible in both arousal and valence detection. This is in contrast with previous evidence provided by literature [8][10], possibly suggesting that the predictive power of EMG decreases when electrodes are placed on the arm rather than over the face, as in previous studies [4][14].

The performance observed in our study is comparable to that reported by the original study [14] (see TABLE IV) thus providing evidence that emotion recognition with low cost noninvasive sensors produces results that are comparable to those achieved with more expensive and invasive sensors. To enable a fair comparison, we remind the reader that we performed a partial replication of the previous study. The peripheral sensor setting of the original study also included consideration of features derived from respiration, blood pressure and eye blinking rate in combination with GSR and EMG. As such, we can directly compare our results only with the original EEG experiment.

Another difference with respect to the original study, is the evaluation setting. In the original study a classifier was trained and tested, for each subject, in a cross-validation setting. Conversely, we use a cross-subject approach, using leave-one-out for evaluation. As such, our models are learnt by considering features extracted from the signals recorded for all the subjects involved. This makes our approach suitable and robust with respect to our target application scenario, patients with impaired mobility and cognition, for which training and fine-tuning classifiers are not feasible.

## 6. Conclusions

We investigated whether it is possible to reliably recognize emotional valence and arousal by using noninvasive low cost sensors. Specifically, EEG, GSR, and EMG sensors were used to collect biometrics. To elicit emotions in subjects involved in our experiment, we used music videos from the DEAP multimodal dataset. We achieved a classification performance comparable to the results reported in the original study that we partially replicate here, even in a cross-subject classification setting that eliminates the need for individual training and tuning of classification models.

Although promising, our results might be further validated with a larger, gender-balanced sample of subjects. In our future research, we also plan to validate the classification performance of our approach by considering a realistic usage scenario, involving patients with impaired cognition and mobility during their medical treatments.

## Acknowledgments

This work is partially supported by the project 'EmoQuest - Investigating the Role of Emotions in Online Question & Answer Sites', funded by the Italian Ministry of Education, University and Research (MIUR) under the program "Scientific Independence of young Researchers" (SIR).